\documentclass[11pt]{article}

\usepackage[margin=1in]{geometry}
\usepackage[T1]{fontenc}
\usepackage[utf8]{inputenc}

\usepackage{newtxtext,newtxmath}

\usepackage{amsmath}
\usepackage{graphicx}
\usepackage{booktabs}
\usepackage{enumitem}

\usepackage[dvipsnames]{xcolor}
\definecolor{TitleColor}{RGB}{28,37,54}
\definecolor{SectionColor}{RGB}{55,78,107}
\definecolor{LinkColor}{RGB}{74,102,136}
\definecolor{RuleColor}{RGB}{185,188,192}

\usepackage[numbers,sort&compress]{natbib}

\usepackage[
    colorlinks=true,
    linkcolor=LinkColor,
    citecolor=LinkColor,
    urlcolor=LinkColor
]{hyperref}

\usepackage[most]{tcolorbox}
\tcbset{
  agentbox/.style={
    breakable,
    enhanced,
    colback=gray!3,
    colframe=RuleColor,
    colbacktitle=gray!12,
    coltitle=black,
    boxrule=0.5pt,
    arc=1.5pt,
    left=8pt, right=8pt, top=6pt, bottom=6pt,
    before skip=14pt, after skip=14pt,
    fonttitle=\normalsize,
    fontupper=\small,
  }
}
\newtcolorbox{understandingbox}[2]{
  agentbox,
  title={\textbf{#1}\\[1pt]{\footnotesize\itshape #2}},
}
\newtcolorbox{recordbox}[1]{
  agentbox,
  title={\textbf{#1}},
}
\newcommand{\recorditem}[1]{\par\smallskip\noindent\textbf{#1:}~}
\newcommand{\recordrule}{\par\medskip\noindent{\color{RuleColor}\hrulefill}\par\smallskip}

\newcommand{\understanding}{\texttt{<understanding>}\xspace}

\usepackage{authblk}

\usepackage{xspace}
\usepackage{setspace}
\usepackage{titlesec}
\titleformat{\section}
  {\large\bfseries}
  {\thesection.}{0.5em}{}
\titleformat{\subsection}
  {\normalsize\bfseries}
  {\thesubsection}{0.5em}{}
\titleformat{\subsubsection}
  {\normalsize\itshape}
  {\thesubsubsection}{0.5em}{}

\title{\textbf{\color{TitleColor} Hypothesis-Driven Autonomous Materials Synthesis\\
with Multimodal LLM Agents}}
\author[1]{Izumi Takahara}
\author[2]{Kazunori Nishio}
\author[3]{Akira Aiba}
\author[2]{Shigeru Kobayashi}
\author[4]{Takao Nakajima}
\author[2]{Taro Hitosugi}
\author[1,*]{Teruyasu Mizoguchi}
\affil[1]{Institute of Industrial Science, The University of Tokyo, Tokyo 153-8505, Japan}
\affil[2]{Department of Chemistry, The University of Tokyo, Tokyo 113-0033, Japan}
\affil[3]{Rigaku Corporation, Tokyo 196-8666, Japan}
\affil[4]{MITSUI KNOWLEDGE INDUSTRY CO., LTD., Tokyo 107-0062, Japan}
\affil[*]{\textit{Corresponding author:} teru@iis.u-tokyo.ac.jp}
\date{}

\begin{document}

\maketitle

\begin{abstract}
\noindent
Self-driving laboratories can explore synthesis conditions autonomously, but their decision-making layer is typically a black-box optimizer, and the output is a set of optimized samples, with the measurements reduced to predefined scalar objectives and the reasons behind success left unarticulated. Here we present SynAgent, a framework in which large language model agents operate an automated experimental system and maintain an explicit, revisable understanding of the synthesis process as the campaign's primary output. Starting with no predefined analysis pipeline, SynAgent adaptively generates analysis skills for newly acquired data and evolves this understanding through multimodal reasoning over experimental data such as X-ray diffraction patterns and electron micrographs. The evolution is guided by a verify--falsify scheme, in which the agent deliberately challenges its own hypotheses by testing conditions predicted to fail as well as those predicted to succeed. In a single campaign of 18 autonomous experiments using LiCoO$_2$ (001) thin-film deposition as a testbed, SynAgent synthesized highly crystalline films and evolved an understanding of how the substrate temperature governs crystallization, discovering an abrupt threshold and a narrow optimal growth window at 650--690~$^\circ$C. These results extend autonomous experimentation beyond optimized samples to testable, human-readable understanding.
\end{abstract}

\section{Introduction}

The development of new functional materials underpins technological progress across diverse fields, from energy to information processing~\cite{Chen2020CriticalReview,Toyao2020CatalysisInformatics,Merchant2023GNoME}. In this endeavor, synthesis plays a central role, requiring researchers to navigate a vast space of processing conditions through repeated cycles of fabrication, characterization, and analysis~\cite{Wang2023ScientificDiscovery}. Because these cycles are slow and labor-intensive, laboratory automation has become a key strategy for accelerating materials research~\cite{Cooper2025Accelerating}. Self-driving laboratories (SDLs), which close the loop between robotic experimentation and machine-learning-based experiment planning, have demonstrated order-of-magnitude accelerations across diverse domains~\cite{Abolhasani2023rise,Tom2024SelfDrivingLabs}, including photocatalyst formulations~\cite{Burger2020mobile}, functional thin films~\cite{Macleod2020self}, inorganic powders~\cite{Szymanski2023alab}, and semiconductor nanocrystals~\cite{Xu2025autonomous}.

In most SDLs, experiment planning is driven by black-box optimization, typically Bayesian optimization (BO), which proposes the next experimental conditions on the basis of the outcomes accumulated so far~\cite{Burger2020mobile,Macleod2020self,Shimizu2020Autonomous,Kobayashi2023Autonomous,Xu2025autonomous}. Black-box optimization is a powerful and well-established strategy for optimizing a predefined objective. When it serves as the decision-making layer of an SDL, however, two limitations arise. First, recent platforms interconnect modular synthesis and measurement instruments and deliver diverse data from every experiment, including diffraction patterns, electron micrographs, spectra, and measured physical properties~\cite{Nishio2025DigitalLaboratory}, yet the optimizer sees only one or a few hand-crafted scalar metrics, leaving most of these data unexploited. Second, black-box optimizers remain largely opaque to researchers, as they can neither explain why they propose a particular condition nor articulate what the accumulated measurements have established about the process and what remains unknown. Indeed, it has been argued that laboratory automation has so far accelerated only the testing of hypotheses, while their generation and refinement remain outside the loop~\cite{Jacobsson2026Hypotheses}.

In parallel, the rapid advance of large language models (LLMs) opens a complementary possibility. Unlike conventional optimizers, LLMs can articulate their reasoning in natural language, opening their decision-making to human inspection. LLMs encode substantial knowledge of inorganic synthesis. Fine-tuned LLMs predict synthesizability and precursors at levels comparable to task-specific machine-learning models~\cite{kim2024large,kim2025explainable}, and even off-the-shelf LLMs recall precursors and processing conditions reported in the literature, in some tasks matching specialized models~\cite{Prein2025language}. Modern LLMs are also multimodal, accepting images and other modalities alongside text, which allows them to interpret heterogeneous experimental data directly. Moreover, the growing reasoning and tool-use capabilities of LLMs have made it possible to deploy them as autonomous agents~\cite{Yao2023ReAct,Shinn2023Reflexion}. Exploiting these agentic capabilities, LLM agents have accelerated materials design in silico~\cite{Jia2024LLMatDesign,Takahara2025MatAgent,Nduma2025Crystalyse,Kim2026Materealize}. They are now beginning to enter physical laboratories as well, where they have assisted tasks ranging from the planning and execution of syntheses to the operation of instruments and the orchestration of research workflows~\cite{Boiko2023Coscientist,Bran2024ChemCrow,Darvish2025ORGANA,mandal2025evaluating,vriza2026operating,shi2026knowledge,Huang2025CASCADE}, and have most recently steered closed-loop solid-state synthesis campaigns~\cite{Fei2026AgenticLLM}.

These advances suggest that LLM agents can serve not merely as optimizers but as participants in the experimental process itself. Serving as such a participant, however, requires capabilities that current frameworks do not yet provide. Many agentic frameworks have been validated only against simulated experiments or computational tasks~\cite{YanguasGil2026ALDOptimization,Cisse2026ClosedLoop,Zou2025ElAgente}, and even the agents that have reached physical laboratories remain largely passive, operating with predefined toolsets and a single data modality. The picture of the system that they build up over a campaign also remains implicit rather than being maintained as an explicit, testable account. Moreover, LLM reasoning is prone to biases, such as anchoring on early exploration choices~\cite{YanguasGil2026ALDOptimization} and overtrust of prior knowledge and plausible-sounding hypotheses~\cite{Cisse2026ClosedLoop}, so that the agent's initially incomplete beliefs about the system may persist unrevised throughout a campaign.

Here we present SynAgent, a multimodal, multi-agent framework in which a main agent plans experiments and interprets their outcomes, supported by subagents that generate analysis skills (Fig.~\ref{fig:overview}). Harnessing the scientific knowledge and reasoning capabilities of LLMs, SynAgent actively tests its own hypotheses through autonomous experiments. To address the challenges outlined above, SynAgent combines two capabilities. First, \textit{adaptive skill generation}: a dedicated subagent generates analysis skills on demand for newly acquired experimental data and registers them for reuse, freeing the framework from predefined analysis pipelines and integrating the diverse data produced during a campaign into the agent's reasoning. Second, \textit{evolving understanding}: the agent generates hypotheses about the relationship between synthesis conditions and outcomes, tests them experimentally, and revises them accordingly. This evolution is driven by a verify--falsify reasoning scheme, in which the agent deliberately tests not only conditions predicted to succeed but also conditions predicted to fail, counteracting the confirmation bias that would otherwise leave incomplete beliefs unchallenged. It is further supported by multimodal reasoning, which integrates X-ray diffraction (XRD) patterns and scanning electron microscopy (SEM) images as complementary evidence. Every decision is accompanied by explicit reasoning that human researchers can trace, interpret, and learn from.

We demonstrate SynAgent on a physical automated laboratory~\cite{Nishio2025DigitalLaboratory} for the synthesis of highly crystalline LiCoO$_2$ (001) thin films, a prototypical layered cathode material~\cite{Mizushima1981LixCoO2} whose cation ordering depends sensitively on deposition conditions~\cite{Antaya1994InSitu,Nishio2019Epitaxial}. This well-characterized dependence makes the system an ideal testbed, allowing us to assess whether the agent can acquire an understanding of the relationship through its own autonomous experiments. We show that SynAgent autonomously generates the analysis skills required for the data at hand and progressively evolves its understanding of the deposition process. It delivers not only promising synthesis conditions but also a clear statement of how substrate temperature governs crystallization, namely an abrupt threshold and a narrow window in which well-ordered films form, written so that a researcher can read it, trace it back to the experiments behind it, and test it further.

\section{Results}

\subsection*{SynAgent framework}

\begin{figure}[!b]
\centering
\includegraphics[width=\textwidth]{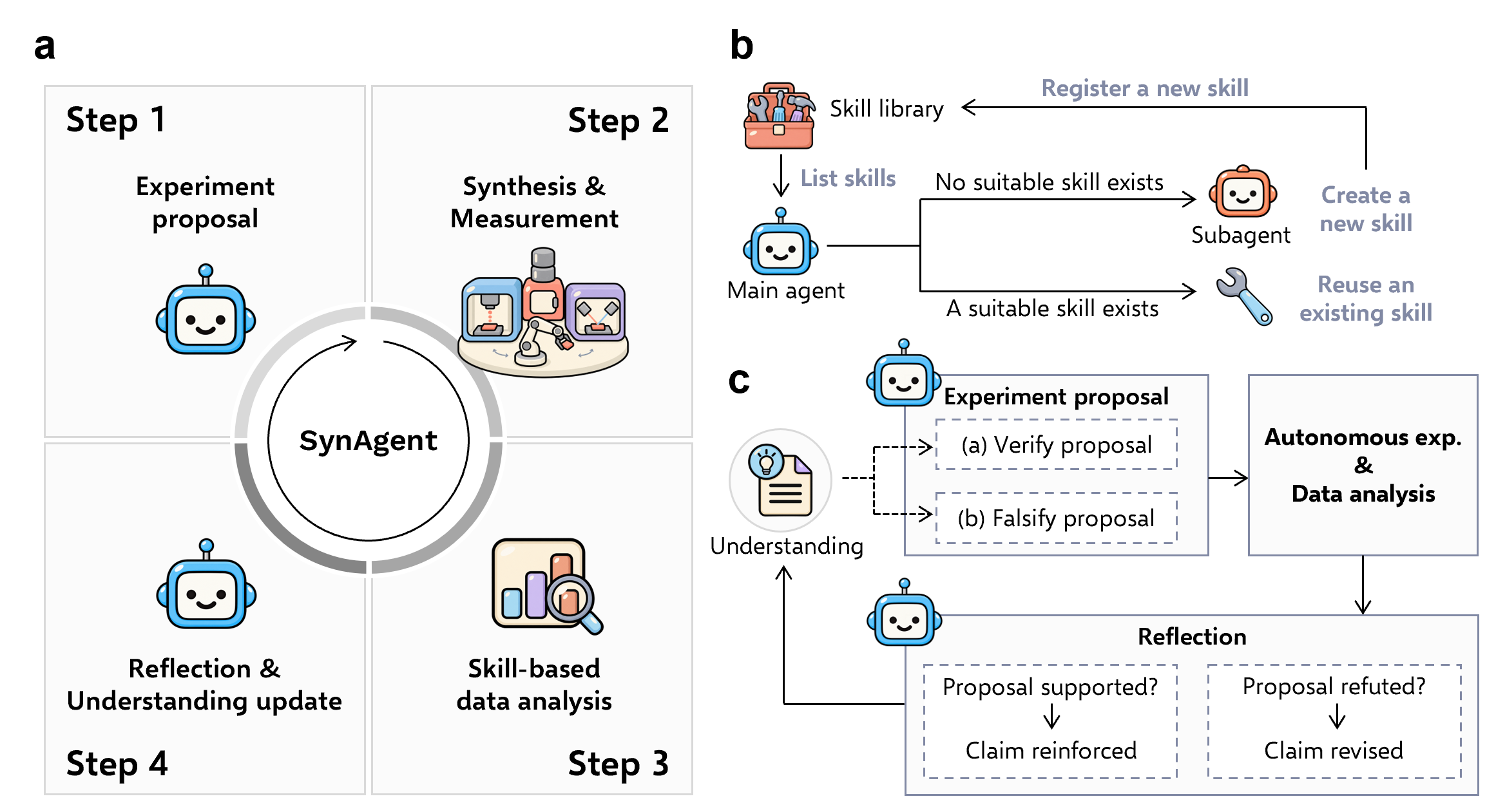}
\caption{\textbf{Overview of the SynAgent framework.}
\textbf{a}, The four-step autonomous loop of experiment proposal (Step 1), synthesis and measurement (Step 2), skill-based data analysis (Step 3), and reflection with understanding update (Step 4), where all reasoning is performed by multimodal LLM agents.
\textbf{b}, Adaptive skill generation: when no suitable skill exists in the skill library, the main agent delegates to a subagent that creates and registers a new one, and the registered skills are reused in later iterations.
\textbf{c}, Evolving understanding through verify--falsify reasoning: each proposal is designed to either verify or falsify the current understanding, and the reflection on the experimental outcome reinforces or revises it, progressively evolving the understanding over the course of the campaign.}
\label{fig:overview}
\end{figure}

Figure~\ref{fig:overview} presents the overall architecture of SynAgent, which couples the scientific knowledge, multimodal reasoning, and code-generation capabilities of LLMs with autonomous experimentation to continually update its understanding of the target system and thereby propose increasingly promising synthesis conditions. SynAgent runs a synthesis campaign as an iterative four-step loop (Fig.~\ref{fig:overview}a): the agent proposes synthesis conditions on the basis of its current understanding (Step 1), synthesis and measurement are executed autonomously under the proposed conditions (Step 2), the acquired data are analyzed with skills, that is, analysis code generated by SynAgent itself (Step 3), and the agent reflects on the outcome and updates its understanding (Step 4).

SynAgent incorporates two mechanisms into this loop. The first, adaptive skill generation (Fig.~\ref{fig:overview}b), frees the framework from the predefined toolsets that constrain current laboratory agents. SynAgent maintains a skill library, a growing collection of the skills generated during the campaign. At the start of a campaign, this library is empty, and no analysis capability is predefined. When newly acquired data call for an analysis that no registered skill covers, the main agent delegates the creation of a new skill to a dedicated subagent. The subagent designs the analysis code around the actual measurement data and registers the validated skill in the library. The registered skills are then reused by the main agent, both when a new measurement arrives and during its reflection on the outcome. The analysis repertoire therefore grows over the course of the campaign rather than being fixed at design time, allowing heterogeneous data such as XRD patterns and SEM images to enter the decision loop as complementary evidence.

The second, the verify--falsify scheme that drives the evolution of understanding (Fig.~\ref{fig:overview}c), follows from the demand that the agent produce an understanding rather than only an optimized sample. If the goal were only optimization, there would be little reason to test conditions predicted to fail. But an understanding is a claim about where the process succeeds and where it fails, and a claim of that kind is confirmed only when its predicted failures are observed. The agent therefore tests conditions it expects to fail as well as those it expects to succeed. Equally important, the same discipline counteracts the susceptibility of LLM reasoning to biases such as anchoring and overconfidence in prior knowledge.

SynAgent maintains its understanding of the target system as an explicit natural-language document, a set of claims about how the synthesis conditions govern the outcome, and updates this document at every cycle of the loop. In each iteration, the agent derives from the current understanding either a verify proposal, targeting conditions under which synthesis of the desired material is predicted to succeed, or a falsify proposal, targeting conditions under which it is predicted to fail, committing in both cases to the hypothesis being tested and to a concrete prediction of the outcome.

After the autonomous experiment and analysis, the agent reflects on whether the outcome matched the committed prediction. This reflection is multimodal, as the agent receives the acquired images themselves as vision input and reasons over them alongside the measured target and the descriptors extracted by the skills. A supported proposal reinforces the underlying claim, whereas a refuted proposal leads the agent to revise the claims now in doubt through abductive reasoning, inferring the most plausible explanation for the unexpected outcome. By deliberately seeking evidence against its own claims, the agent prevents an initially incomplete picture of the system from persisting uncorrected, and the progressively refined understanding in turn becomes the basis for the subsequent proposals.
\subsection*{Synthesis of LiCoO$_2$ (001) thin films}

\begin{figure}[!p]
\centering
\includegraphics[width=\textwidth]{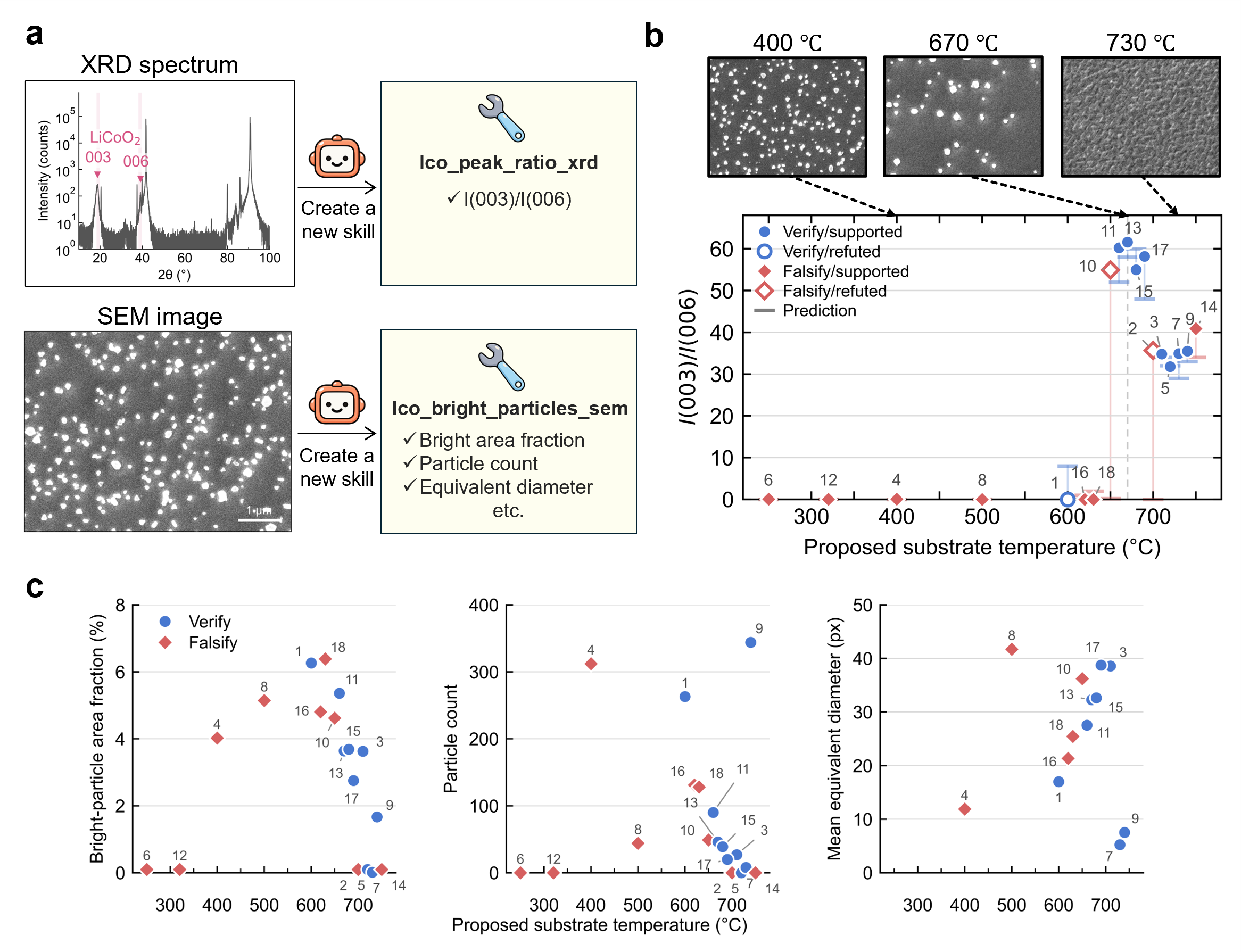}
\caption{\textbf{Behavior of SynAgent in the LiCoO$_2$ synthesis campaign.}
\textbf{a}, Analysis skills generated by the agent from the data of the first experiment (600~$^\circ$C). For the X-ray diffraction (XRD) pattern, the agent generated a skill (\texttt{lco\_peak\_ratio\_xrd}) that searches for the LiCoO$_2$ 003 and 006 reflections within windows around their reference positions (pink bands) and computes their intensity ratio, $I(003)/I(006)$. Pink triangles mark the maxima selected by the skill. For the scanning electron microscopy (SEM) image, the agent identified bright particles scattered over the film surface, features visible to the eye but not part of any predefined analysis, and generated a skill (\texttt{lco\_bright\_particles\_sem}) that quantifies the surface morphology, including the bright-particle area fraction, the particle count, and the equivalent particle diameter.
\textbf{b}, $I(003)/I(006)$ as a function of the proposed substrate temperature. Numbers denote iterations, blue circles denote verify proposals, and red diamonds denote falsify proposals. Filled and open symbols indicate that the prediction was supported and refuted, respectively. Horizontal bars indicate the ratios predicted by the agent, and are connected to the corresponding measured values. The vertical dashed line marks 670~$^\circ$C (iteration 13), which yielded the highest ratio. Insets show SEM images of the films grown at 400, 670, and 730~$^\circ$C.
\textbf{c}, The generated SEM skill returns three morphological descriptors, namely the bright-particle area fraction, the particle count, and the mean equivalent particle diameter (in pixels, with 5~nm per pixel), as functions of the proposed substrate temperature. Iterations with no detected particles are omitted from the diameter panel.}
\label{fig:licoo2}
\end{figure}

To demonstrate the framework, we applied SynAgent to the synthesis of highly crystalline LiCoO$_2$ (001) thin films on a physical automated laboratory~\cite{Nishio2025DigitalLaboratory}, with all agents driven by GPT-5.5, a multimodal LLM developed by OpenAI~\cite{OpenAI2026GPT55}. In this system, high crystallinity refers to a well-ordered layered structure, in which the Li and Co ions occupy alternating cation layers. We considered the growth of LiCoO$_2$ (001) on Al$_2$O$_3$ (0001) substrates, in which the crystallinity is controlled by the substrate temperature. The agent was asked to use the intensity ratio of the 003 and 006 reflections in the XRD pattern, $I(003)/I(006)$, as a measure of the crystallinity of LiCoO$_2$~\cite{Nishio2019Epitaxial,Nishio2025DigitalLaboratory}, and to propose substrate temperatures that maximize this ratio. The agent was therefore required to observe how the proposed temperatures relate to the crystallinity of the resulting films and to evolve its understanding of this relationship over the course of the campaign. In this campaign, the agent alternated between verify and falsify proposals at successive iterations, beginning with a verify proposal (Fig.~\ref{fig:overview}c).

Figure~\ref{fig:licoo2} summarizes the behavior of SynAgent in this campaign, showing the analysis skills it generated and the outcomes of all the experiments analyzed by these skills. Figure~\ref{fig:licoo2}a shows the analysis skills that the agent generated from the data measured in the first experiment, for which it proposed a substrate temperature of 600~$^\circ$C. To compute the objective $I(003)/I(006)$ from the XRD pattern, the agent generated an analysis skill that locates the 003 and 006 reflections and extracts their intensity ratio. During the reflection on this experiment, the agent recognized bright particles dispersed on the film surface in the SEM image, and requested another skill tailored to this specific morphology, quantifying the bright-particle area fraction, the particle count, and the equivalent particle diameter, together with statistics of the background texture. These descriptors served as complementary evidence for interpreting the synthesis outcomes, and the values they returned agreed with manual analysis of the same data.

Figure~\ref{fig:licoo2}b summarizes the measured $I(003)/I(006)$ of all the experiments as a function of the proposed substrate temperature, together with the agent's predictions. Each experiment is classified by the mode of the proposal, verify or falsify, and by whether the measured ratio supported or refuted the prediction committed to at the proposal stage. Over the campaign, the agent proposed and executed 18 experiments, of which 15 predictions were judged as supported and 3 as refuted (Supplementary Table~1). The measured ratios reveal a sharply structured temperature dependence. The films grown at 250--630~$^\circ$C showed essentially no ordering, with ratios below 0.1, whereas ordered growth set in abruptly at 650~$^\circ$C, with a narrow high-quality window at 650--690~$^\circ$C and a plateau of ratios of 32--41 at 700--750~$^\circ$C. The highest crystallinity of the campaign, $I(003)/I(006) = 61.6$, was obtained at 670~$^\circ$C. The surface morphology also varied markedly with the temperature, as illustrated by the SEM images of the films grown at 400, 670, and 730~$^\circ$C (insets in Fig.~\ref{fig:licoo2}b).

Figure~\ref{fig:licoo2}c shows the morphological descriptors extracted from the SEM images, which allowed the agent to distinguish between films that are indistinguishable by XRD alone. Both regimes below 650~$^\circ$C show ratios below 0.1 and are identical as far as the objective is concerned. The films grown at 250--320~$^\circ$C showed nearly featureless surfaces at the imaging resolution, with bright-particle area fractions of about 0.1\%, which the agent tentatively interpreted as continuous films formed under mobility-limited conditions. In contrast, the films grown at 400--630~$^\circ$C contained bright particles covering 4--6\% of the surface. The agent interpreted these features as possible signatures of incomplete crystallization or surface segregation, although their composition and origin could not be determined from the SEM images alone. On the high-temperature plateau, the SEM images showed continuous films with roughened or faceted textures but few visible segregated particles, which the agent used to attribute the lower ratios of this region to subtler causes such as defects, texture, strain, or mild stoichiometry shifts, while noting that the composition of the faceted features would require additional characterization to determine.

In proposing each substrate temperature, the agent drew on all of this evidence, namely the measured $I(003)/I(006)$, the SEM images themselves, and the morphological descriptors extracted from them. The raw images also contributed observations that were not captured by the numerical descriptors, such as the faceted shapes of the bright surface features. These observations informed the agent's tentative hypotheses about possible growth and failure modes, although they did not establish the composition or phase identity of the observed features. All of these analyses were performed with the two skills generated at the first iteration, which were reused without modification throughout the campaign. Taken together, these results demonstrate that SynAgent, integrating the XRD objective with the complementary SEM evidence into an understanding evolved from its own observations, narrowed the search to the 650--690~$^\circ$C window and located the best condition at 670~$^\circ$C in a single autonomous campaign, while generating testable hypotheses, in its own words, about why the films outside that window exhibited lower ordering.

\subsection*{Evolving understanding through verify--falsify reasoning}

\begin{figure}[!b]
\centering
\includegraphics[width=\textwidth]{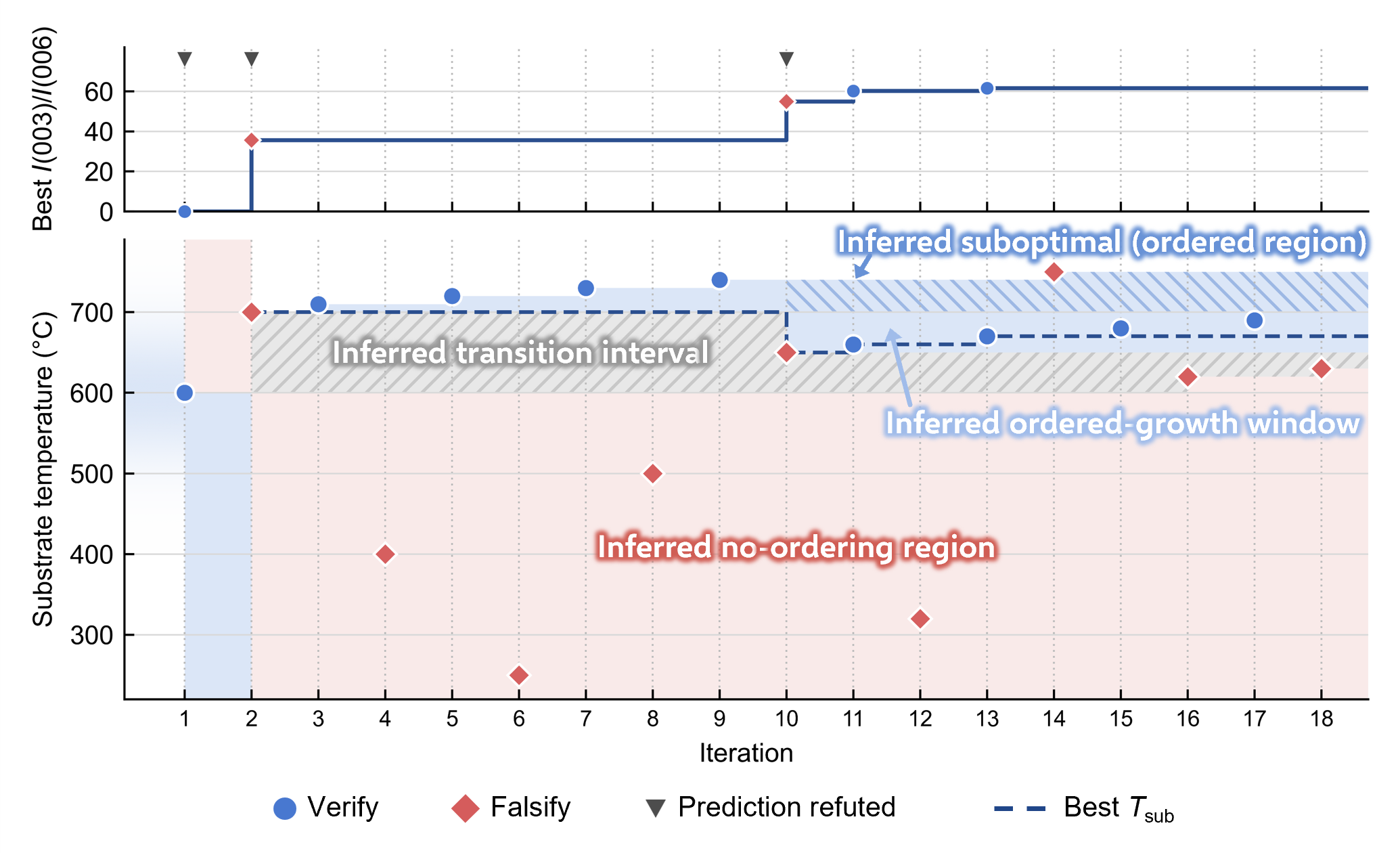}
\caption{\textbf{Evolution of the agent's understanding over the campaign.}
Top, the best $I(003)/I(006)$ ratio obtained up to each iteration. Gray triangles mark the iterations at which the agent's prediction was refuted (iterations 1, 2, and 10).
Bottom, the substrate temperatures proposed at each iteration, shown as blue circles for verify proposals and red diamonds for falsify proposals, overlaid on the temperature regions stated in the agent's evolving understanding. Each region extends from the iteration after which it was stated until it was revised by a later experiment, and the shading before iteration 1 schematically represents the prior expectation that 600~$^\circ$C would be near optimal. The dashed line traces the substrate temperature that the agent regarded as best at each stage.}
\label{fig:evolution}
\end{figure}

Figure~\ref{fig:evolution} traces how this picture was established, following the campaign iteration by iteration. The top panel shows the best $I(003)/I(006)$ obtained up to each iteration, and the bottom panel shows the proposed substrate temperatures overlaid on the temperature regions stated in the agent's evolving understanding, which culminated in a partition of the temperature axis into four regions. The full text of the understanding at the key stages of the campaign is reproduced in Supplementary Note~1. The first verify proposal tested the prior expectation that 600~$^\circ$C would be near optimal, and failed with $I(003)/I(006) = 0.001$ (Supplementary Note~2). Reflecting on the diffraction pattern and the particle-covered surface observed by SEM, the agent hypothesized that the failure might be associated with incomplete crystallization, morphological inhomogeneity, or possible secondary-phase segregation, and revised its understanding to place the optimum below 600~$^\circ$C. The falsify proposal of the next iteration probed 700~$^\circ$C as a condition predicted to fail under this revised picture, but instead yielded a well-ordered film with a ratio of 35.6. This single refutation overturned the revised picture, and the agent reorganized its understanding into a temperature-activated crystallization threshold, above which ordering sets in abruptly, and placed the transition somewhere between 600 and 700~$^\circ$C. The subsequent verify proposals extended the newly found ordered growth from 700~$^\circ$C up to 740~$^\circ$C, establishing a plateau of similar ratios, while the interleaved falsify proposals at 250--500~$^\circ$C confirmed the absence of ordering at low temperatures.

The decisive step came at iteration 10. The falsify proposal probed 650~$^\circ$C, the unsampled gap between the failed 600~$^\circ$C film and the successful 700~$^\circ$C film, with a predicted ratio of 0.2. The measured ratio of 54.9 was instead the best obtained so far, refuting the claim that the crystallization threshold lay close to 700~$^\circ$C (Supplementary Note~2). The agent narrowed the transition interval to 600--650~$^\circ$C, and the following verify proposals rapidly refined the newly opened window: 660~$^\circ$C gave 60.2 and 670~$^\circ$C gave 61.6, the highest crystallinity of the campaign, while 680 and 690~$^\circ$C gave 55.0 and 58.2. The remaining falsify proposals sharpened the boundaries of this picture. The film grown at 750~$^\circ$C gave 40.9, supporting the claim that the high-temperature plateau is robust but suboptimal, and the films grown at 620 and 630~$^\circ$C failed completely, bracketing the abrupt transition between 630 and 650~$^\circ$C. The SEM image of the 630~$^\circ$C film still showed abundant faceted surface features, which the agent regarded as being consistent with a morphologically inhomogeneous pre-threshold growth regime. However, SEM alone could not determine whether these features represented secondary phases, compositional segregation, or LiCoO$_2$ crystallites with a distinct morphology. By the end of the campaign, the agent had thus bracketed the transition to a 20~$^\circ$C interval, and resolving it further would require a finer temperature grid.

Each of the three refuted predictions triggered a major revision of the understanding. The final understanding delineated the four regions shown in Fig.~\ref{fig:evolution}: a no-ordering region below 630~$^\circ$C, a transition interval between 630 and 650~$^\circ$C, an ordered-growth window at 650--690~$^\circ$C with the optimum near 670~$^\circ$C, and a robust but suboptimal ordered region at 700--750~$^\circ$C.

These results illustrate how the verify--falsify scheme drove the campaign. The preconception inherited from prior knowledge, that 600~$^\circ$C would be near optimal, was overturned at the first iteration, when the prediction committed by the verify proposal was refuted. The two largest advances that followed, the discovery of ordered growth at 700~$^\circ$C and of the superior window at 650~$^\circ$C, were both made by falsify proposals. This is not incidental. Proposals whose outcomes match their predictions refine the current picture, as when the falsify proposals at 620 and 630~$^\circ$C bracketed the transition. It is when a predicted failure does not occur that the agent learns something it could not have anticipated, and the experiment is informative in proportion to the confidence with which the failure was predicted. Each refutation was converted into a revision of the understanding. The verify--falsify scheme thus corrects the preconceptions that the LLM brings into a campaign and drives its understanding toward a faithful picture of the system, which in this case guided the agent to the optimal growth window.

\section{Discussion}

In this work, we proposed SynAgent, a multi-agent framework that connects LLM agents to an automated experimental system and drives autonomous materials synthesis campaigns while generating its own analysis tools and actively testing its own hypotheses. Demonstrated on a physical automated laboratory, SynAgent synthesized and characterized 18 LiCoO$_2$ (001) films without human intervention, wrote and registered the two analysis skills it needed at the first iteration, and evolved an explicit, human-readable understanding of the deposition process that culminated in the four-region map of Fig.~\ref{fig:evolution} and in the optimal growth window at 650--690~$^\circ$C.

The product of the campaign is therefore twofold. Beyond the synthesis conditions themselves, SynAgent delivers an interpretable account of the process, a set of testable mechanistic hypotheses that state where ordered growth occurs and why, each linked to the experiments that support or constrain it. In this respect the framework complements rather than replaces BO. The LiCoO$_2$ task, the temperature grid, and the objective were the same as those of the BO-driven campaign reported previously on this platform~\cite{Nishio2025DigitalLaboratory}, and SynAgent likewise arrived at the optimal condition. The difference lies in what remains at the end of the campaign: BO leaves the optimum together with the data that produced it, whereas SynAgent additionally leaves a document that partitions the temperature axis into four regions, locates an abrupt threshold between 630 and 650~$^\circ$C, and states that the 700--750~$^\circ$C plateau is robust but suboptimal, with each claim tied to the experiments that supported or refuted it. This account, however, was possible only because the agent could decide what features to extract and quantify from the acquired measurements. The bright particles on the low-temperature films were not part of any predefined analysis pipeline. The agent saw them in the SEM image, judged them informative, and obtained the means to quantify them. An agent restricted to a predefined toolset can reason only over quantities that someone anticipated, and an unanticipated feature is precisely what an evolving understanding needs. The two skills generated at the first iteration were reused unchanged for the remaining 17 experiments, indicating that what the agent produced were not ad hoc scripts but analysis tools. The multimodal evidence proved essential to this account, as the SEM-derived descriptors separated the featureless low-temperature films from the particle-covered ones, a distinction invisible to the scalar objective, and thereby shaped the mechanistic hypotheses that the agent subsequently refined and tested.

The campaign also illustrates why deliberate falsification matters for LLM-driven experimentation. LLM agents are prone to overtrust their prior knowledge and to anchor to their own early successes, failure modes reported for simulated campaigns~\cite{YanguasGil2026ALDOptimization,Cisse2026ClosedLoop}. An agent rewarded only for confirming its expectations would have had little reason to revisit the region between the failed 600~$^\circ$C film and the successful high-temperature plateau. In our campaign, the falsify proposals were precisely what forced such revisits, and their refutations produced the two largest advances. By committing to a concrete prediction before each experiment and revising its claims through abduction when the prediction failed, the agent replaced an initially biased picture with an understanding grounded in the observed data, converting surprises into structured updates.

Several avenues remain for extending the framework. The present demonstration involved a single controllable variable and a single materials system, and applying the same scheme to higher-dimensional condition spaces, where concise and testable understanding documents would be even more valuable, is a natural next step. In this campaign, the verify and falsify modes were simply alternated. Exploring more adaptive switching strategies, for example letting the agent choose the mode according to the maturity of its current understanding, may make the evolution of understanding still more efficient. Each condition was, moreover, visited only once, and the small differences among the best ratios in the 650--690~$^\circ$C window may lie within run-to-run variation. Repeated experiments would quantify this variation and further test the claims of the understanding. A controlled comparison with BO would likewise require re-running both decision layers under matched budgets, which the sequential nature of the physical experiments precluded here. Finally, the correctness of the generated skills is currently verified only within the scope of their self-tests, and the judgments of the campaign, from the supported and refuted verdicts to the abductive explanations, rest on the reasoning of the LLM itself. Since prior knowledge can enter a campaign through the design of its analysis tools, human-in-the-loop or agent-in-the-loop approaches, in which the generated skills and the recorded judgments are reviewed by human researchers or by specialized validation agents, would be an effective complement.

In conclusion, these results position LLM agents not merely as optimizers of laboratory workflows but as experimental partners that decide how newly acquired measurements should be analyzed, and build, articulate, and stress-test scientific understanding. As self-driving laboratories continue to standardize and interconnect instruments~\cite{Nishio2025DigitalLaboratory}, such agents will be able to draw on ever richer characterization modalities and to carry the understanding gained in one campaign into the next. We anticipate that autonomous experiments of this kind will deliver not only better materials but also a deeper understanding of the underlying materials science.

\section{Methods}

\subsection*{SynAgent framework}
SynAgent is implemented as a set of LLM agents built on the OpenAI Agents SDK~\cite{OpenAI2025AgentsSDK}, with OpenAI GPT-5.5~\cite{OpenAI2026GPT55} as the underlying model for all agents. A campaign is declared in a configuration file that specifies an overview of the experiment, the target variable, the controllable variables with their allowed grids, and the fixed experimental conditions. Each iteration consists of three LLM turns around the autonomous experiment. In the proposal turn, the agent receives the campaign declaration, the observed pairs of conditions and target values, and its current understanding of the system, which is maintained as a natural-language document denoted \understanding, and returns, under the mode assigned by the schedule, a structured proposal consisting of the hypothesis under test, the proposed condition chosen from the allowed grid, a concrete predicted target value, and the reasoning behind them. After the proposed experiment has been executed and the measurement analyzed, the reflection turn judges the outcome, and the understanding turn rewrites \understanding. All turns run inside a single persistent session stored on disk, so that the agent retains the conversational context of the entire campaign. The SEM images of the three most recent iterations are replayed to the model as vision input, and older images are replaced by the textual morphology descriptions that the agent itself recorded at the corresponding iterations. Every prompt and response is logged per iteration, and all proposals, measured targets, skill outputs, and judgments are appended to a persistent history file, making each decision traceable.

\subsection*{Adaptive skill generation}
A skill consists of a metadata document (\texttt{SKILL.md}) and an analysis module (\texttt{analyze.py}). The metadata describe the technique, material, instrument, metric, and applicability of the skill, and the module implements a single function that reads a measurement file in the standardized MaiML format~\cite{Nishio2025DigitalLaboratory,JAIMA_MaiML} and returns a structured analysis result. When a measurement arrives, a dispatcher presents the metadata of the registered skills to the LLM, which either selects a matching skill or signals that a new one is needed. A new skill is generated by a dedicated agent equipped with a code-execution tool. The generator agent inspects the actual MaiML file and the available reference files, and iteratively writes and runs candidate analysis code against the measured data within a fixed tool-call budget. When the MaiML file references an image, the image is also attached to the generator as vision input, so that the analysis is designed for the features actually present in the data. Before registration, the generated module is executed on the measurement at hand, and the skill enters the library only when this self-test passes. On failure, the error and the previous code are fed back for regeneration, with up to three attempts. During reflection, the main agent can execute any registered skill on the measurements of the current iteration through a \texttt{use\_skill} tool and can request at most one new skill per iteration through a \texttt{create\_skill} tool, specifying what to quantify and from which measurement. The prompt instructs the agent to prefer existing skills and to create a new one only when no registered skill covers the needed analysis. The skill that computes the optimization target is pinned by name in the campaign configuration. It is generated once, when the first measurement arrives, and its definition remains fixed for the rest of the campaign. In the LiCoO$_2$ campaign, the XRD skill was generated through this route, the SEM skill was created through a \texttt{create\_skill} request during the first reflection, and no further skills were created in the remaining iterations.

\subsection*{Verify--falsify reasoning scheme}
The proposal mode alternates between verify and falsify, starting from verify, and is implemented as an interchangeable strategy so that other schedules can be substituted. In each proposal turn, the agent identifies the single claim of \understanding to be tested, chooses a condition that tests this claim under the current mode, and commits to a concrete numerical prediction of the target. In the reflection turn, the agent receives the measured target value and the SEM image and judges the prediction as supported or refuted following a mode-dependent procedure. In verify mode, the prediction is judged as supported when the measurement and the prediction both indicate success, and in falsify mode, when both indicate failure. No numerical tolerance is imposed on the difference between the predicted and measured values. The numerical prediction serves as a concrete commitment, and the agent itself judges whether the outcome falls into the predicted category. Every judgment is accompanied by a written reasoning that compares the measured and predicted values in the light of the hypothesis and the SEM observation. When the prediction is supported, the agent states which claim of \understanding is now more strongly backed by the experiment, without extrapolating beyond the tested regime. When the prediction is refuted, the agent performs abduction, naming the claim now in doubt and proposing the most plausible alternative explanation for the observed outcome. The understanding turn then rewrites \understanding to incorporate the judgment, and the updated document becomes the basis of the next proposal.

\subsection*{Autonomous synthesis and characterization}
All experiments were performed on dLab, a digital laboratory that physically interconnects modular synthesis and measurement instruments~\cite{Nishio2025DigitalLaboratory}. The campaign task, including the temperature grid and the objective, follows the autonomous demonstration reported previously on the same platform, in which the next condition was selected by BO. In the present work, this decision-making layer is replaced by SynAgent. (001)-oriented LiCoO$_2$ thin films were grown on Al$_2$O$_3$ (0001) substrates by RF magnetron sputtering using a sintered Li$_{1.2}$CoO$_x$ target. The RF power was fixed at 80~W, the deposition time at 1.5~h, and the total gas pressure at 0.50~Pa, with argon and oxygen partial pressures of 0.45 and 0.05~Pa. The substrate temperature during deposition was the only controllable variable and was chosen by the agent from 200 to 750~$^\circ$C in steps of 10~$^\circ$C. The campaign was run for a preset budget of 18 iterations. XRD patterns were measured with a Rigaku SmartLab XE diffractometer over a $2\theta$ range of 10--120$^\circ$ with a step of 0.01$^\circ$. Surface morphologies were observed with a JEOL JSM-IT700HR scanning electron microscope at a pixel size of 5.0~nm. All measurement data were delivered to the agent as MaiML files, and the SEM data associated with each XRD measurement were linked automatically.

\section*{Acknowledgments}
This work was supported by JST ACT-X (grant no. JPMJAX24DB), JST BOOST (grant no. JPMJBS2418), JST CREST (grant no. JPMJCR2204), the MEXT Data-Driven Materials Research and Development Project (grant no. JPMP1122712807) and JSPS KAKENHI (grant no. 24K01599).

\section*{Data and Code Availability}
The data and code for this study are available on request to the corresponding author.

\section*{Declaration of Generative AI and AI-Assisted Technologies}
During the preparation of this work, the authors used Claude Code (Anthropic) and ChatGPT (OpenAI) to assist in developing the software used in the experiments and in preparing the manuscript. The authors reviewed and edited the output as needed and take full responsibility for the content of the published article.

\bibliographystyle{unsrt}
\bibliography{references}

\clearpage
\setcounter{table}{0}
\setcounter{figure}{0}
\renewcommand{\thetable}{S\arabic{table}}
\renewcommand{\thefigure}{S\arabic{figure}}

\phantomsection
\addcontentsline{toc}{section}{Supplementary Information}
\begin{center}
{\Large\bfseries\color{TitleColor} Supplementary Information for:\\[4pt]
Hypothesis-Driven Autonomous Materials Synthesis with Multimodal LLM Agents\par}
\end{center}
\vspace{1em}

\noindent
This Supplementary Information contains Supplementary Table~1, summarizing all iterations of the autonomous LiCoO$_2$ synthesis campaign, Supplementary Note~1, reproducing the agent's understanding document at key stages of the campaign, and Supplementary Note~2, reproducing the full proposal and reflection records of the three iterations whose predictions were refuted.

\begin{table}[!htb]
\centering
\caption{\textbf{Summary of the autonomous LiCoO$_2$ synthesis campaign.}
For each iteration, the table lists the proposal mode, the proposed substrate temperature $T_{\mathrm{sub}}$, the predicted and measured $I(003)/I(006)$, and the judgment of the reflection turn. In verify mode, the prediction is judged as supported when the measurement and the prediction both indicate successful ordered growth. In falsify mode, it is judged as supported when both indicate failure. No numerical tolerance was imposed on the difference between the predicted and measured values, and the agent itself made the judgment. All 18 judgments were decided by whether the outcome fell into the predicted category: the three refuted judgments correspond to reversals between success and failure, whereas deviations within the same category, up to a factor of eight at 400~$^\circ$C, were judged as supported.}
\label{tab:campaign}
\begin{tabular}{cccccc}
\toprule
 & & & \multicolumn{2}{c}{$I(003)/I(006)$} & \\
\cmidrule(lr){4-5}
Iteration & Mode & $T_{\mathrm{sub}}$ ($^\circ$C) & Predicted & Measured & Judgment \\
\midrule
1 & verify & 600 & 8.0 & 0.001 & refuted \\
2 & falsify & 700 & 0.001 & 35.6 & refuted \\
3 & verify & 710 & 34.0 & 34.8 & supported \\
4 & falsify & 400 & 0.010 & 0.081 & supported \\
5 & verify & 720 & 32.0 & 31.8 & supported \\
6 & falsify & 250 & 0.005 & 0.056 & supported \\
7 & verify & 730 & 29.0 & 34.9 & supported \\
8 & falsify & 500 & 0.050 & 0.001 & supported \\
9 & verify & 740 & 33.0 & 35.5 & supported \\
10 & falsify & 650 & 0.200 & 54.9 & refuted \\
11 & verify & 660 & 52.0 & 60.2 & supported \\
12 & falsify & 320 & 0.050 & 0.043 & supported \\
13 & verify & 670 & 58.0 & 61.6 & supported \\
14 & falsify & 750 & 34.0 & 40.9 & supported \\
15 & verify & 680 & 60.0 & 55.0 & supported \\
16 & falsify & 620 & 1.0 & 0.001 & supported \\
17 & verify & 690 & 48.0 & 58.2 & supported \\
18 & falsify & 630 & 2.0 & 0.001 & supported \\
\bottomrule
\end{tabular}
\end{table}

\clearpage

\section*{Supplementary Note 1: Evolution of the understanding document}

\begin{sloppypar}
SynAgent maintains its evolving understanding of the target synthesis process as an explicit natural-language document, denoted \understanding{} in the main text, and rewrites this document at the end of every iteration.
\end{sloppypar}
\begin{sloppypar} This note reproduces the document verbatim at four key stages of the LiCoO$_2$ campaign: after iteration 1, in which the initial expectation that 600~$^\circ$C would be near optimal was refuted, after iteration 2, in which the falsify proposal discovered ordered growth at 700~$^\circ$C, after iteration 10, in which the falsify proposal discovered the superior growth window at 650~$^\circ$C, and after the final iteration 18, in which the understanding reached the four-region map described in the main text. The header of each box summarizes the proposal and the outcome of the corresponding iteration. Bold face follows the emphasis placed by the agent in the original document, and only typographic conversion, such as the typesetting of chemical formulae, was applied, leaving the wording unchanged.
\end{sloppypar}

\begin{understandingbox}{Understanding after iteration 1}{verify proposal at 600~$^\circ$C, predicted $I(003)/I(006)$ = 8.0, measured 0.001, judged refuted}
At \textbf{600~$^\circ$C}, LiCoO$_2$ growth under the fixed sputtering conditions fails to produce ordered layered (001) crystallinity: the measured \textbf{$I(003)/I(006)$ = 0.001} is essentially zero, contradicting the initial assumption that 600~$^\circ$C would be near-optimal.

The SEM morphology supports this failure: the surface is mostly fine-textured/dark but contains many bright isolated particles/islands, with an estimated bright-particle area fraction of $\sim$6\%. This suggests inhomogeneous growth, likely from \textbf{secondary-phase segregation, Li/Co oxide particulate formation, or islanding}, rather than a continuous well-ordered layered LiCoO$_2$ film.

The most plausible mechanism is that, at 600~$^\circ$C and the given oxygen pressure, the film becomes chemically or morphologically unstable: \textbf{Li volatility/off-stoichiometry} and/or \textbf{secondary-phase formation} may dominate over the beneficial effect of increased adatom mobility. Thus, the previous claim that intermediate-high temperature improves ordering up to around 600~$^\circ$C is now doubtful.

Current working hypothesis: the optimal substrate temperature, if it exists, is likely \textbf{below 600~$^\circ$C}, where Li loss and secondary-phase segregation may be reduced while still allowing enough adatom mobility for layered ordering.
\end{understandingbox}

\begin{understandingbox}{Understanding after iteration 2}{falsify proposal at 700~$^\circ$C, predicted $I(003)/I(006)$ = 0.001, measured 35.6, judged refuted}
The prior claim that the optimum is likely \textbf{below 600~$^\circ$C} is now \textbf{rejected}. A 700~$^\circ$C film gave \textbf{$I(003)/I(006)$ = 35.64}, while 600~$^\circ$C gave \textbf{0.001}, so the temperature response is strongly non-monotonic and cannot be explained by simple high-temperature degradation.

The prior observation that \textbf{600~$^\circ$C is a failed growth condition} is reinforced. Its particle-rich SEM morphology and near-zero XRD ratio still indicate poor layered LiCoO$_2$ formation, likely from a local instability such as phase segregation or incomplete transformation.

New working hypothesis: \textbf{700~$^\circ$C crosses a crystallization/mobility threshold} that enables strong layered LiCoO$_2$ (001) ordering on Al$_2$O$_3$ under the fixed sputtering/O$_2$ conditions. In this regime, the dominant mechanism appears to be improved adatom mobility and crystallization kinetics, producing a continuous fine-textured film with essentially no bright secondary-phase particles in SEM.

The 600~$^\circ$C failure is now best interpreted not as evidence that all higher temperatures cause Li loss, but as a possible \textbf{intermediate-temperature instability window} where mobility or reaction kinetics are insufficient for complete layered ordering, allowing segregated particles/islands to form.

Open questions:
\begin{itemize}[leftmargin=1.4em,itemsep=0.15em,topsep=0.2em]
  \item Is 700~$^\circ$C near the optimum, or does performance improve further at 710--750~$^\circ$C?
  \item Does Li volatility/off-stoichiometry eventually dominate at still higher temperature?
  \item Where is the sharp transition between failed 600~$^\circ$C growth and successful 700~$^\circ$C growth?
\end{itemize}

Discriminating evidence would be XRD/SEM at temperatures between 600 and 700~$^\circ$C and above 700~$^\circ$C: a smooth increase would support a crystallization-threshold mechanism, while a peak followed by degradation would indicate onset of Li-loss-driven instability.
\end{understandingbox}

\begin{understandingbox}{Understanding after iteration 10}{falsify proposal at 650~$^\circ$C, predicted $I(003)/I(006)$ = 0.200, measured 54.9, judged refuted}
The prior claim that the crystallization threshold lies close to \textbf{700~$^\circ$C} is \textbf{rejected}. The \textbf{650~$^\circ$C} sample produced \textbf{$I(003)/I(006)$ = 54.92}, the best result so far and much higher than the 700--740~$^\circ$C values ($\sim$32--36). Thus, 650~$^\circ$C is not below the ordering threshold; it is now the leading candidate for the optimum.

The updated working hypothesis is that the useful ordered-growth window begins somewhere between \textbf{600 and 650~$^\circ$C}, with a sharp transition from the failed 600~$^\circ$C condition to highly ordered growth by 650~$^\circ$C. In this window, the dominant mechanism is likely that substrate temperature is high enough to enable \textbf{Li/Co/O adatom mobility and layered-phase crystallization}, while still low enough to avoid some high-temperature roughening, defect formation, or Li volatility that may slightly limit the 700--740~$^\circ$C plateau.

The 600~$^\circ$C failure is now best viewed as a \textbf{local pre-threshold instability or incomplete-crystallization regime}, not as representative of all temperatures below 700~$^\circ$C. At 600~$^\circ$C, mobility/reaction kinetics may be just insufficient for complete layered ordering, allowing segregated particles or poorly ordered phases to dominate.

The high-temperature success window remains reinforced but is no longer assumed optimal: \textbf{700--740~$^\circ$C} consistently gives strong ordering, but not as high as 650~$^\circ$C. This suggests that excess thermal energy may introduce more defects, roughening/coarsening, or slight stoichiometric deviation even though the layered phase remains dominant.

Morphology interpretation is further refined: the 650~$^\circ$C SEM shows sparse bright triangular/faceted features ($\sim$4.6\% bright area fraction), yet the XRD ratio is excellent. Therefore, faceted bright features can coexist with very high layered ordering and are not automatically destructive secondary phases. Their composition remains unknown.

Current working map:
\begin{itemize}[leftmargin=1.4em,itemsep=0.15em,topsep=0.2em]
  \item \textbf{250~$^\circ$C:} continuous but likely amorphous/nanocrystalline; poor ordering from insufficient mobility.
  \item \textbf{400--600~$^\circ$C:} poor/failure regime with incomplete crystallization and particle/island segregation.
  \item \textbf{650~$^\circ$C:} best observed condition; likely near optimal balance of mobility, layered ordering, and limited high-temperature degradation.
  \item \textbf{700--740~$^\circ$C:} robust but lower plateau of strong layered ordering.
\end{itemize}

Open questions:
\begin{itemize}[leftmargin=1.4em,itemsep=0.15em,topsep=0.2em]
  \item Where exactly between \textbf{600 and 650~$^\circ$C} does the sharp transition occur?
  \item Is the true optimum at \textbf{650~$^\circ$C} or nearby, e.g. 630--680~$^\circ$C?
  \item Does 750~$^\circ$C remain high-performing or begin to show upper-temperature degradation?
  \item Are the bright triangular/faceted features at 650 and 740~$^\circ$C LiCoO$_2$ crystallites, secondary phases, or surface-segregated material? Secondary-peak inspection or compositional analysis would discriminate these possibilities.
\end{itemize}
\end{understandingbox}

\begin{understandingbox}{Understanding after iteration 18}{falsify proposal at 630~$^\circ$C, predicted $I(003)/I(006)$ = 2.0, measured 0.001, judged supported}
The lower crystallization threshold is now constrained more tightly: \textbf{630~$^\circ$C} still fails completely with \textbf{$I(003)/I(006)$ = 0.001}, while \textbf{650~$^\circ$C} is excellent at \textbf{54.92}. Therefore, the abrupt transition to high-quality layered LiCoO$_2$ ordering occurs between \textbf{630 and 650~$^\circ$C}.

The claim that \textbf{630~$^\circ$C remains in the pre-threshold incomplete-crystallization/segregation regime} is reinforced. SEM at 630~$^\circ$C shows many bright triangular/faceted particles or islands on a dark continuous background: \textbf{128 bright objects}, $\sim$\textbf{6.4\% bright area fraction}, typical diameter $\sim$\textbf{25 px}. Because this morphology coincides with near-zero XRD ratio, these features likely represent segregated islands/secondary crystallites or incomplete phase formation rather than coherent layered LiCoO$_2$ (001) growth.

The dominant mechanism remains a \textbf{temperature-activated crystallization threshold}. Below the threshold, Li/Co/O mobility and layered-phase reaction kinetics are insufficient, so the film forms a poorly ordered matrix plus particles/islands. Above the threshold, the matrix crystallizes into strongly layered LiCoO$_2$, and similar faceted surface features can become benign rather than destructive.

Current working map:
\begin{itemize}[leftmargin=1.4em,itemsep=0.15em,topsep=0.2em]
  \item \textbf{250--320~$^\circ$C:} mobility-limited amorphous/nanocrystalline continuous films; very poor ordering.
  \item \textbf{400--630~$^\circ$C:} incomplete crystallization/segregation regime; near-zero ordering with particles/islands.
  \item \textbf{650--690~$^\circ$C:} high-quality window; optimum near \textbf{660--670~$^\circ$C}, with \textbf{670~$^\circ$C} currently best.
  \item \textbf{700--750~$^\circ$C:} good but suboptimal high-temperature growth, likely limited by subtle defects, texture, strain, or mild stoichiometry shifts.
\end{itemize}

Remaining key open question: the transition lies between \textbf{630 and 650~$^\circ$C}; testing \textbf{640~$^\circ$C} would locate whether the switch is very sharp near 640--650~$^\circ$C.
\end{understandingbox}

\clearpage

\section*{Supplementary Note 2: Proposal and reflection records of the refuted iterations}

Each iteration of the campaign is recorded as a structured proposal, committed before the experiment, and a structured reflection, produced after the measurement. This note reproduces these records verbatim for the three iterations whose predictions were refuted (iterations 1, 2, and 10), which triggered the major revisions of the understanding described in the main text. The proposal record lists the mode, the proposed condition, the committed prediction, the hypothesis under test, and the reasoning behind the proposal. The reflection record lists the measured target, the judgment, the reasoning behind the judgment, the observation of the SEM image, and, because the prediction was refuted, the abduction that names the claim now in doubt and proposes the most plausible alternative explanation.

\begin{recordbox}{Proposal and reflection record of iteration 1}
\recorditem{Mode} verify

\recorditem{Proposed condition} substrate temperature 600~$^\circ$C

\recorditem{Predicted target} $I(003)/I(006)$ = 8.0

\recorditem{Hypothesis} LiCoO$_2$ layered ordering and (001) crystallinity are maximized at an intermediate-high substrate temperature where adatom mobility is sufficient for cation ordering without excessive Li volatility or secondary-phase formation.

\recorditem{Reasoning} With no prior observations, I test the common sputtered LiCoO$_2$ mechanism that increasing substrate temperature improves surface diffusion and layered ordering up to an optimum, while too high a temperature can degrade stoichiometry via Li loss. A substrate temperature of 600~$^\circ$C is a plausible near-optimal verification point for crystalline layered LiCoO$_2$ on sapphire under fixed oxygen pressure. The expected $I(003)/I(006)$ value is set to 8.0 as a concrete prediction for a successful, well-ordered film under this hypothesis.

\recordrule

\recorditem{Measured target} $I(003)/I(006)$ = 0.001

\recorditem{Judgment} refuted

\recorditem{Reflection} The verification prediction was a successful high $I(003)/I(006)$ ratio of \textbf{8.0} at 600~$^\circ$C, but the measured value was \textbf{0.001}, essentially indicating no useful 003/006 layered LiCoO$_2$ diffraction ratio. This is qualitatively opposite to the hypothesis that 600~$^\circ$C would provide sufficient adatom mobility for well-ordered (001) LiCoO$_2$ without damaging stoichiometry.

\recorditem{SEM observation} The SEM image shows a mostly dark, fine-textured/continuous background with many bright isolated particles or islands distributed across the surface. Automated morphology analysis estimates about \textbf{6.3\% bright-particle area fraction} with \textbf{263 bright objects} in the image and typical equivalent diameters around \textbf{17 px}. This particle-rich, inhomogeneous morphology is consistent with secondary-phase particles, Li/Co oxide segregation, droplets, or Volmer-Weber-like islanding rather than a smooth epitaxial layered film, supporting the very low XRD ratio.

\recorditem{Abduction} The claim now in doubt is that \textbf{600~$^\circ$C lies in the optimal intermediate-high temperature regime where mobility improves layered ordering without causing degradation}. The evidence instead points to the alternative that, under the fixed sputtering/O$_2$ conditions, \textbf{600~$^\circ$C is already too high or chemically unfavorable}, likely promoting Li volatility, off-stoichiometry, secondary-phase segregation, or particulate/island formation. These mechanisms would suppress coherent LiCoO$_2$ (001) crystallinity and explain both the near-zero $I(003)/I(006)$ ratio and the bright particle-rich SEM morphology.
\end{recordbox}

\begin{recordbox}{Proposal and reflection record of iteration 2}
\recorditem{Mode} falsify

\recorditem{Proposed condition} substrate temperature 700~$^\circ$C

\recorditem{Predicted target} $I(003)/I(006)$ = 0.001

\recorditem{Hypothesis} If the current interpretation is correct that high-temperature growth above the useful window promotes Li volatility/off-stoichiometry and secondary-phase or island formation, then an even hotter deposition temperature should also fail to produce ordered layered LiCoO$_2$ (001) diffraction.

\recorditem{Reasoning} The only sampled point, 600~$^\circ$C, failed strongly, and the mechanism inferred was high-temperature chemical/morphological instability from Li loss and secondary-phase segregation. In falsify mode, 700~$^\circ$C deliberately probes a broad under-sampled high-temperature region away from the suspected below-600~$^\circ$C optimum; if the mechanism is correct, this condition should underperform rather than recover high layered ordering. I therefore predict an essentially failed $I(003)/I(006)$ ratio of 0.001, comparable to the 600~$^\circ$C failure.

\recordrule

\recorditem{Measured target} $I(003)/I(006)$ = 35.6

\recorditem{Judgment} refuted

\recorditem{Reflection} The falsify prediction was that 700~$^\circ$C would fail similarly to 600~$^\circ$C, with expected $I(003)/I(006)$ = \textbf{0.001}, because high temperature was assumed to worsen Li volatility/off-stoichiometry and secondary-phase formation. Instead, the measured ratio is \textbf{35.63830235}, an unexpectedly strong success and qualitatively opposite to the predicted failure. Therefore the high-temperature-failure hypothesis is refuted for 700~$^\circ$C.

\recorditem{SEM observation} The SEM image is largely continuous and featureless/fine-textured, without the bright particle-rich morphology seen at 600~$^\circ$C. Quantitative SEM analysis found essentially no bright secondary-phase particles: bright area fraction $\sim$0.001 and particle count 0. The background is rough/noisy but uniformly covered, consistent with a more continuous film and reduced visible secondary-phase segregation, matching the high XRD ratio.

\recorditem{Abduction} The claim now in doubt is that \textbf{temperatures above 600~$^\circ$C necessarily promote Li loss/off-stoichiometry and secondary-phase/island formation under these fixed sputtering conditions}. The new evidence points to an alternative: \textbf{600~$^\circ$C may be a local poor-growth regime}, while at \textbf{700~$^\circ$C} adatom mobility, crystallization kinetics, and/or phase formation cross a threshold that enables well-oriented layered LiCoO$_2$ before detrimental Li loss dominates. The disappearance of bright particles suggests that the segregated/particulate morphology at 600~$^\circ$C was not simply monotonic high-temperature degradation, but may reflect incomplete crystallization, transient phase separation, or a narrow instability window.
\end{recordbox}

\begin{recordbox}{Proposal and reflection record of iteration 10}
\recorditem{Mode} falsify

\recorditem{Proposed condition} substrate temperature 650~$^\circ$C

\recorditem{Predicted target} $I(003)/I(006)$ = 0.200

\recorditem{Hypothesis} The sharp crystallization threshold has not been reached well below 700~$^\circ$C; at 650~$^\circ$C adatom mobility and layered-phase reaction kinetics should still be insufficient for complete LiCoO$_2$ (001) ordering, leading to underperforming or segregated/incompletely crystallized films rather than the high-ratio 700--740~$^\circ$C regime.

\recorditem{Reasoning} Falsify mode should probe an under-sampled region away from the verified 700--740~$^\circ$C success cluster. The most informative unsampled gap is between the failed 600~$^\circ$C point and the successful 700~$^\circ$C point. Choosing 650~$^\circ$C tests whether the transition occurs only close to 700~$^\circ$C; under the current threshold model, 650~$^\circ$C is expected to remain below the mobility/crystallization threshold and fail or strongly underperform. I predict $I(003)/I(006)$ = 0.2, allowing slight improvement over 600~$^\circ$C but still far below a successful film.

\recordrule

\recorditem{Measured target} $I(003)/I(006)$ = 54.9

\recorditem{Judgment} refuted

\recorditem{Reflection} The falsify prediction was that 650~$^\circ$C would remain below the crystallization threshold and underperform, with expected \textbf{$I(003)/I(006)$ = 0.2}. Instead, the measured value is \textbf{54.92220144}, the highest result so far and far above the 700--740~$^\circ$C plateau. This is a surprise success and refutes the claim that 650~$^\circ$C is below the threshold for strong layered LiCoO$_2$ ordering.

\recorditem{SEM observation} The SEM shows a continuous rough/fine-textured background with sparse but prominent bright triangular/faceted crystallites or islands. Quantitative analysis found \textbf{49 bright particles}, \textbf{$\sim$4.6\% bright area fraction}, and typical equivalent diameter $\sim$\textbf{36 px}. Despite these faceted features, the XRD ratio is exceptionally high, so the morphology is more consistent with coexistence of strong layered LiCoO$_2$ ordering with faceted crystalline growth or minor surface segregation, rather than destructive secondary-phase-dominated failure.

\recorditem{Abduction} The prior claim now in doubt is that the crystallization threshold lies close to 700~$^\circ$C and that 650~$^\circ$C should still be mobility-limited. The evidence instead points to the alternative working hypothesis that \textbf{650~$^\circ$C is within, or possibly near the optimum of, the ordered LiCoO$_2$ growth window}. A plausible mechanism is that 650~$^\circ$C provides enough adatom mobility for layered ordering while reducing high-temperature roughening, Li volatility, or excess defect formation compared with higher temperatures. The very poor 600~$^\circ$C result may therefore represent a sharp local transition or instability just below the ordered-growth window rather than evidence for a broad 600--700~$^\circ$C failure regime.
\end{recordbox}

\end{document}